\documentclass[]{pasj01}

\usepackage{natbib}
\newcommand{\nar}{New Astronomy Reviews}

\begin{document} 
\Received{2016/02/08}
\Accepted{2016/03/15}
\Published{}

\title{A Radio Detection Survey of Narrow-Line Seyfert~1 Galaxies using Very-Long-Baseline Interferometry at 22~GHz}

%
%

\author{
Akihiro \textsc{Doi}\altaffilmark{1,2}, 
Tomoaki \textsc{Oyama}\altaffilmark{3}, 
Yusuke \textsc{Kono}\altaffilmark{4}, 
Aya \textsc{Yamauchi}\altaffilmark{3}, 
Syunsaku, \textsc{Suzuki}\altaffilmark{4},
Naoko \textsc{Matsumoto}\altaffilmark{5,4}, and 
Fumie \textsc{Tazaki}\altaffilmark{4}
}

\altaffiltext{1}{The Institute of Space and Astronautical Science, Japan Aerospace Exploration Agency, 3-1-1 Yoshinodai, Sagamihara, Kanagawa 229-8510}
\altaffiltext{2}{Department of Space and Astronautical Science, SOKENDAI (The Graduate University for Advanced Studies), 3-1-1 Yoshinodai, Sagamihara, Kanagawa 229-8510}
\altaffiltext{3}{Mizusawa VERA Observatory, National Astronomical Observatory of Japan, 2-12 Hoshigaoka, Mizusawa, Oshu, Iwate 023-0861}
\altaffiltext{4}{Mizusawa VLBI Observatory, NAOJ, 2-21-1 Osawa, Mitaka, Tokyo 181-8588}
\altaffiltext{5}{The Research Institute for Time Studies, Yamaguchi University, 1677-1 Yoshida, Yamaguchi, Yamaguchi 753-8511}


\maketitle

\begin{abstract}
We conducted a high-sensitivity radio detection survey for forty narrow-line Seyfert~1 (NLS1) galaxies using very-long-baseline interferometry (VLBI) at 22~GHz through phase-referencing long-time integration and using a newly developing recorder with a data rate of 8~Gbps, which is a candidate of the next generation VLBI data recording systems for the Japanese VLBI Network.  The baseline sensitivity was typically a few mJy.  The observations resulted in a detection rate of 12/40 for our radio-selected NLS1 sample; 11 out of the detected 12 NLS1s showed inverted radio spectra between 1.4 and 22~GHz on the basis of the Very Large Array flux densities and the VLBI detections.  These high fractions suggest that a compact radio core with a high brightness temperature is frequently associated with NLS1 nuclei.  
On the other hand, at least half of the sample indicated apparently steep spectra even with the limited VLBI sensitivity.  Both the inverted and steep spectrum radio sources are included in the NLS1 population.  
\end{abstract}

\section{Introduction}  
The radio natures of narrow-line Seyfert~1 (NLS1s) galaxies potentially provide us key understandings of outflowing mechanisms in the growing phase of active galactic nuclei (AGNs).  NLS1s as a class are thought to be fed at high mass accretion rates onto relatively small-mass black holes, potentially connecting between stellar-mass and supermassive black hole systems in the mass function.    
The first systematic studies in radio bands by interferometric observations were carried out at arc-second resolutions using the Very Large Array~(VLA), and indicated little difference between NLS1s and Seyfert galaxies: similar radio luminosities, steep radio spectra, and scarcely resolved radio morphology suggest the presence of weak nonthermal jets as a radio emitting source \citep{Ulvestad:1995,Moran:2000}, although several optical/X-ray properties are clearly different.  
NLS1s were thought to be radio-quiet objects as a class with only a few exceptions of known radio-loud objects (e.g., \citealt{Grupe:2000,Oshlack:2001,Siebert:1999}); subsequent systematic studies based on large database, such as the VLA Faint Images of the Radio Sky at Twenty-centimeters (FIRST; \citealt{Becker:1995}) and the NRAO VLA Sky Survey (NVSS; \citealt{Condon:1998}) in radio and the Sloan Digital Sky Survey (SDSS) in optical, have revealed a lower fraction of radio-loud objects ($\sim7$\%; \citealt{Komossa:2006}, see also \citealt{Zhou:2006, Whalen:2006, Yuan:2008}) compared to that of broad-line Seyfert galaxies.

Recently, gamma-ray detections by {\it Fermi Gamma-Ray Space Telescope} toward six NLS1s galaxies with high significance \citep{Abdo:2009a,Abdo:2009,DAmmando:2012,DAmmando:2015a} have offered a new population of gamma-ray emitting AGNs other than blazars and radio galaxies.  
The radio observations at high angular resolutions using very-long-baseline interferometry~(VLBI) for the first discovered $\gamma$-ray emitting NLS1 SDSS~J094857.31+002225.4 (PMN~J0948+0022) had revealed the presences of a one-sided pc-scale jet and a rapidly variable, very high brightness radio core showing an inverted spectrum at the nucleus \citep{Doi:2006}.  
PMN~J0948+0022 also shows a core-dominant structure with two-sided kpc-scale radio emissions \citep{Doi:2012}.  
The combination of these radio/gamma-ray properties is reminiscent of blazars, which are characterized by Doppler-beaming on relativistic jets viewed from pole-on typically at parsec~(pc) scales and by decelerated components at from kpc to Mpc scales.         
VLBI observations of the other several gamma-ray emitting NLS1s also indicate the presence of beamed jets at pc scales \citep{DAmmando:2012, DAmmando:2013a, Doi:2011a, Doi:2013a, Wajima:2014}; some of them also show two-sided kpc-scale radio structures \citep{Anton:2008,Doi:2011a,Doi:2012}.   
However, the radio jets are mildly relativistic and their powers are comparable to the least energetic blazars, on the basis of single-dish monitoring \citep{Angelakis:2015}.

On the other hand, VLBI observations of radio-loud NLS1s including gamma-ray detections/non-detections have revealed that the radio-loud aspect is possibly attributed not only to a beaming effect but also an intrinsically large radio power.      
Intrinsic radio powers that are corrected for the beaming effect using estimated Doppler factors suggest that radio-loud NLS1s include both intrinsically radio-loud and intrinsically radio-quiet cases \citep{Doi:2011a,Doi:2012}.  
An extended pc-scale jet showing a steep spectrum dominates a total radio power, which suggests being not so relativistic and/or sufficiently inclined with respect to our line of sight for the jet \citep{Gu:2010,Gu:2015}.  
Seven out of known ten NLS1s with kpc-scale radio structures (\citealt{Richards:2015}, see also \citealt{Gliozzi:2010,Doi:2012,Doi:2015}) are not still gamma-ray detected.  This situation suggests that jet powers may be intrinsically large sufficient to escape the core regions of host galaxies, but not so beamed because of relatively large viewing angles.        
For several nearby radio-quiet NLS1s, VLBI observations have also revealed pc-scale jet-like structures showing steep spectra, which are presumably generated through the same process as that of radio-loud NLS1s, but probably with an only difference in power or viewing angle for nonthermal jets \citep{Giroletti:2005,Doi:2013a,Doi:2015}.   
Thus, at least a fraction of NLS1 nuclei has an ability to generate nonthermal jets, which have a wide range of radio powers; in some cases, the signatures of beaming effect on relativistic jets viewed from pole-on are observed.  

However, a limited number of NLS1s has still been investigated at milli-arcsecond (mas) resolutions so far, because NLS1s are relatively weak radio sources in comparison with the other AGN classes.   
Systematic VLBI observations have been started by the other authors \citep{Gu:2015}, whose targets are sixteen radio-loud NLS1s selected from several parent samples with 105~sources; fourteen sources were detected at 5 and 6.7~GHz, which were relatively low frequencies and may tend to be biased to the contribution of an extended jet component.     
The present paper reports the result of a new VLBI detection survey as a systematic study for NLS1s in the radio band at a higher frequency (22~GHz), where an inner jet component is potentially focused on.  
Our observation was planned before the publication of \citet{Gu:2015}.  Our sample selection  was relatively similar to theirs; the differences are (1)~the combination of parent samples, (2)~including radio-quiet NLS1s as well, and (3)~at a higher observing frequency.     
Section~\ref{section:sample} presents the sample selection.  In Sections~\ref{section:observation} and \ref{section:reduction}, VLBI observations and the procedures of data reductions are described.  These results are presented and its implications are briefly discussed in Section~\ref{section:result}.

\section{Sample}\label{section:sample}
We selected NLS1 radio sources by position-matching between the catalog of the Very Large Array~(VLA) 1.4-GHz Faint Images of the Radio Sky at Twenty cm~(FIRST) survey \citep{Becker:1995} and source lists in the following NLS1 studies: (1)~64~sources in \citet{Veron-Cetty:2001a}, (2)~2011~sources in \citet{Zhou:2006}, (3)~23~sources in \citet{Yuan:2008}, and (4)~62~sources in \citet{Whalen:2006}.  We found several sources that were counted redundantly; the number of uniquely selected sources resulted in 233~sources (734--0.8~mJy~beam$^{-1}$).  We clipped these sources into 41~sources with an intensity higher than 10~mJy~beam$^{-1}$.  The NLS1 radio source catalogue that is a target list in the present study is listed in Table~\ref{table:sample}.

The sample selection criteria are similar to those of the previous systematic VLBI study by \citet{Gu:2015}, which was also based on the radio selected samples with flux densities of $>10$~mJy in the FIRST.  The parent samples were restricted to only radio-loud objects from \citet{Komossa:2006,Zhou:2006,Yuan:2008,Foschini:2011}, which were slightly different from ours including radio-quiet ones as well.  Six radio-quiet objects are included in our sample ($\log{RL}<1$, see Col.~(7) in Table~\ref{table:sample}).

\section{Observation}\label{section:observation}
The observations are performed as a part of the experimental observations to test a developing OCTAVE-DAS (Data Acquisition System, \citealt{Oyama:2012}).  It is a candidate of the next generation data recording system for OCTAVE~(Optically Connected Array for VLBI Exploration, \citealt{Kono:2012}), JVN~(Japanese VLBI Network, \citealt{Fujisawa:2016}) and other VLBI arrays.   
  
OCTAVE-DAS consists of three key components.  The first is a high speed analog-to-digital~(A/D) converter at a sampling rate of 8~gigabits per~second~(Gbps) of 3~bit quantization. It is called OCTAD~(OCTave A/D converter).  It is able to sample signals at a radio frequency~(RF) directly.  In addition, it has digital base-band converter~(DBBC) functions for the VLBI Global Observing System~(VGOS) observation.  The second key component is a media converter between single 10~GigE~(Gigabit Ethernet) port and four VSI-H~(Vlbi Standard Interface-Hardware) I/O ports.  It is called OCTAVIA~(OCTave Vlbi Interface Adapter) or OCTAVIA2 depending on its version.  The last key component is a recorder.  It is able to record the data stream at the rate from 4.5~Gbps up to 32~Gbps.  It is called OCTADISK~(OCTave DISK recorder), OCTADISK2 or VSREC (Vlbi Software Recorder) depending on its version.  OCTAVE-DAS is being developed at the Mizusawa VLBI observatory, a branch of the National Astronomical Observatory of Japan.  

The test observations using OCTAVE-DAS were conducted on April~21 and 28, 2014 using four radio telescopes of the VLBI Exploration of Radio Astrometry project~(VERA; \citealt{Kobayashi:2003}); Mizusawa, Ogasawara, Iriki, and Ishigaki stations participated in the experiment.  The polarization is the left-circular-polarization.  VERA has a dual-beam receiving system. The dual beams are called A- and B-beam.  For the dual beam observation in this paper, A-beam was used for the target sources. B-beam was used for the reference sources.  

The received radio signals of dual beams are converted to IF frequency, respectively.  The bandwidth is 512~MHz.  The IF signals are sampled by two A/D converters~(ADS-1000s).  The sampling rate is 1.024~GHz.  The quantization bit number is two.  The data streams are recorded by OCTADISK.  The aggregated bit rate is 4.096~Gbps.

In addition to this current VERA observation system, we installed OCTAVE-DAS in order to expand the bandwidth.  We also installed the analog signal converters with wider bandwidth.  The new system was installed on A-beam for this test observation to increase the sensitivity of target sources.  The received signal of A-beam was divided to two signals for the current and the new system.  The signal for the new systems is divided to four IF channels of which bandwidth is 512~MHz.  The starting frequencies of the four IFs are 21.459, 21.971, 22.483, and 22.995~GHz.  The four IF signals are A/D converted by ADS-3000+ which was developed by NICT~(National Institute of Information and Communications Technology) instead of OCTAD for the observation in this paper.  The sampled data stream was recorded by VSREC.  The aggregated recording rate is 8.192~Gbps.

The dual beam systems with the current VERA system and OCTAVE-DAS can work simultaneously.  We utilized ``dual-beam phase referencing'' in order to increase coherent integration time by elimination of the atmospheric phase fluctuation of the targets with the phases of the calibrators.  For several targets (SBS~0846+513, SDSS~J103727.44+003635.5, B3~1441+476, [HB89]~1519-065, [HB89]~1546+353), proper calibrators cannot be found within a separation angle of 2.1 degrees. We used a nodding style phase-referencing.  A-beam pointing was switched between a target and a reference calibrator.  On-source tracking was used for PMN~J0948+0022, which was expected to be detected without phase-referencing.  SDSS~J124634.64+023809.0 was mis-allocated in our observation.  The number of observed targets resulted in 40.   

The A/D converter ADS1000 at Ishigaki station was unlocked to a reference signal during both of the two observations unfortunately.  The solutions of phase reference from the B-beam were unavailable.

\section{Data reduction}\label{section:reduction}
The correlations were processed with the software correlator OCTACOR2~(OCTAve CORrelator), which was developed at the Mizusawa VLBI observatory and NICT \citep{Oyama:2012}.  
The correlated data were integrated every one second.  The frequency channel number per an IF channel is 512.  

Data reduction procedures were performed using the Astronomical Image Processing System~(AIPS; \citealt{Greisen:2003}).  Amplitude calibration using a priori gain values together with the system noise temperatures measured during the observations were applied. The calibration accuracy had not been evaluated in this new system, probably less than 20\%, which was inferred from the comparison of results of known strong sources in the data.  
The delay differences among the four IF channels were calibrated using the fringe-fitting solutions of a bright calibrator (''manual-pcal'').  This solution allowed us to adopt a small fringe-finding window in subsequent fringe fitting procedures.  

For dual-beam phase-referencing, the fringe-fitting solution in phase for the reference calibrator was obtained with the B-beam data at first.  Next, we applied the solution to the four channels of the A-beam.
Calibrators for five targets were not detected.  
The difference between the center frequencies of two beams is~256 MHz~($\sim1$\% with respect to the radio frequency).  In the case of residual phase variation of $\sim100$~deg~(typically observed during VERA dual-beam observations) after a fringe-rate removal, the coherence loss is expected to be only 0.02\% when we apply fringe-phase solutions to data at a $\sim1$\%-different frequency.  
With the dual beam phase referencing, we were able to extend the fringe-fitting solution interval to 720~seconds.  The interval is equivalent to a scan duration time of the antenna schedule.  
As a result, six out of 29~targets were detected at signal-to-noise ratios~($SNRs$) higher than 3 in this observing mode.  

We also performed fringe-fitting on the five target that observed in the manner of nodding style phase-referencing with a cycle of 40~seconds.  The solution interval was 720~seconds with a net accumulation of 240~seconds on a target.  As a result, two out of the five targets were detected at $SNR>3$ in this observing mode.  

PMN~J0948+0022, which was observed without phase-referencing, was detected with a solution interval of 120~sec.  Fringe findings for five targets whose calibrators were not detected in B-beam were also attempted with a solution interval of 120~sec.  As a result,  three out of the five were detected at $SNR>3$.


\begin{table*}
\caption{Samples and Results of Observations}
\footnotesize{
\begin{tabular}{llllrcrllllr}
\hline\hline
NED name & $z$ & RA. (J2000) & Dec. (J2000) & $I_\mathrm{1.4GHz}^\mathrm{FIRST}$ & Sample & $\log{RL}$ & Ref.  & Mode & Calibrator & $F_\mathrm{VLBI}$ \\
(1) & (2) & (3) & (4) & (5) & (6) & (7) & (8) & (9) & (10) & (11) \\
\hline
2MASX J03474022+0105143        & 0.031 & 03 47 40.195 & +01 05 14.25 & 38.85 & a & 0.99 &  & 2B & J0352+0238 & $<7$ \\
FBQS J0713+3820                & 0.123 & 07 13 40.291 & +38 20 40.08 & 10.43 & b & 0.33 &  & 2B & J0709+3737 & $<7$ \\
2MASS J07440228+5149175 & 0.46 & 07 44 02.242 & +51 49 17.48 & 11.89 & b & 1.62 &  & 2B & J0733+5022 & $<7$ \\
FBQS J075800.0+392029          & 0.096 & 07 58 00.047 & +39 20 29.09 & 10.8 & b & $-0.19$ &  & 2B & J0752+3730 & $<7$ \\
SDSS J081432.11+560956.6       & 0.510361 & 08 14 32.135 & +56 09 56.55 & 69.18 & c & 2.53 & e & 2B & J0824+5552 & 117 \\
SBS 0846+513                   & 0.584701 & 08 49 57.990 & +51 08 28.83 & 344.09 & c & 3.16 & f & ND & J0905+4850 & 454 \\
SDSS J085001.16+462600.5       & 0.524316 & 08 50 01.171 & +46 26 00.41 & 20.9 & c & 2.23 & e & 2B & J0847+4609 & $<7$ \\
SDSS J090227.16+044309.5       & 0.533025 & 09 02 27.152 & +04 43 09.40 & 152.59 & c & 3.02 & e & 2B & J0901+0448 & 244 \\
PMN J0948+0022 & 0.585102 & 09 48 57.295 & +00 22 25.60 & 107.53 & c & 2.55 & ghi & 1B & \ldots & 501 \\
MRK 1239                       & 0.019927 & 09 52 19.099 & -01 36 43.63 & 58.64 & a & 1.33 & jk & 2B & J0945-0153 & $<7$ \\
SDSS J095317.09+283601.4       & 0.65891 & 09 53 17.106 & +28 36 01.63 & 44.58 & c & 2.71 & e & 2B & J1001+2911 & $<7$ \\
SDSS J103123.73+423439.3       & 0.377167 & 10 31 23.728 & +42 34 39.40 & 16.57 & c & 2.34 &  & 2B & J1038+4244 & $<7$ \\
KUG 1031+398                   & 0.042443 & 10 34 38.599 & +39 38 28.17 & 23.98 & a & 1.67 &  & 2B & J1033+4116 & $<7$ \\
SDSS J103727.44+003635.5       & 0.595596 & 10 37 27.454 & +00 36 35.76 & 27.23 & c & 2.66 & e & ND & J1048+0055 & $<12$ \\
$[$HB89$]$ 1044+476                & 0.79902 & 10 47 32.654 & +47 25 32.24 & 734.02 & c & 3.87 & e & 2B & J1051+4644 & $<7$ \\
SDSS J111005.03+365336.3       & 0.62995 & 11 10 05.034 & +36 53 36.12 & 18.62 & c & 2.97 & e & 2B & J1104+3812 & $<7$ \\
2MASX J11193404+5335181        & 0.105975 & 11 19 34.026 & +53 35 18.45 & 15.49 & d & 1.96 &  & 2B & J1120+5404 & $<7$ \\
SDSS J113824.54+365327.1       & 0.356743 & 11 38 24.545 & +36 53 26.99 & 12.54 & c & 2.34 & e & 2B & J1130+3815 & $<7$ \\
2MASX J11404788+4622046        & 0.11439 & 11 40 47.897 & +46 22 04.82 & 78.85 & b & 1.36 &  & 2B & J1138+4745 & $<7$ \\
FBQS J114654.2+323652          & 0.4658 & 11 46 54.298 & +32 36 52.24 & 14.67 & c & 2.11 &  & 2B & J1152+3307 & 104 \\
FBQS J1151+3822                & 0.334575 & 11 51 17.757 & +38 22 21.75 & 10.93 & b & 0.54 &  & 2B & J1146+3958 & $<7$ \\
2MASX J12022678-0129155        & 0.150694 & 12 02 26.806 & -01 29 15.54 & 11.37 & d & 1.49 &  & 2B & J1207-0106 & $<7$ \\
NGC 4051                       & 0.002336 & 12 03 09.594 & +44 31 52.52 & 12.3 & a & 0.47 & lj & 2B & J1203+4510$\dagger$ & $<17$ \\
NGC 4253                       & 0.012929 & 12 18 26.516 & +29 48 46.52 & 38.16 & a & 1.19 &  & 2B & J1217+3007 & $<7$ \\
SDSS J123852.12+394227.8       & 0.622668 & 12 38 52.147 & +39 42 27.59 & 10.36 & c & 2.23 &  & 2B & J1242+3751 & $<7$ \\
SDSS J124634.64+023809.0       & 0.362629 & 12 46 34.683 & +02 38 09.02 & 37.05 & c & 2.38 & e & \ldots & J1250+0216 & \ldots \\
MRK 0783                       & 0.0672 & 13 02 58.925 & +16 24 27.49 & 18.53 & a & 1.36 & j & 2B & J1300+141B & $<9$ \\
SDSS J130522.74+511640.2       & 0.787552 & 13 05 22.746 & +51 16 39.55 & 83.87 & c & 2.34 & e & 2B & J1259+5140 & $<9$ \\
FBQS J1421+2824                & 0.539978 & 14 21 14.075 & +28 24 52.23 & 46.79 & b & 2.14 & e & 2B & J1419+2706 & 117 \\
SDSS J143509.49+313147.8       & 0.502218 & 14 35 09.523 & +31 31 48.30 & 39.26 & c & 2.87 &  & 2B & J1435+3012$\dagger$ & $<23$ \\
B3 1441+476                    & 0.705472 & 14 43 18.578 & +47 25 56.53 & 164.79 & c & 3.07 & e & ND & J1452+4522 & $<16$ \\
$[$HB89$]$ 1502+036                & 0.407882 & 15 05 06.467 & +03 26 30.83 & 365.39 & c & 3.19 & mn & 2B & J1458+0416$\dagger$ & 697 \\
$[$HB89$]$ 1519-065                & 0.083121 & 15 22 28.758 & -06 44 41.83 & 11.8 & a & 0.56$\dagger\dagger$ &  & ND & J1510-0543 & $<16$ \\
$[$HB89$]$ 1546+353                & 0.479014 & 15 48 17.924 & +35 11 28.37 & 140.94 & c & 2.84 & eo & ND & J1602+3326 & 22 \\
IRAS  15462-0450               & 0.099792 & 15 48 56.806 & -04 59 34.26 & 10.68 & a & 1.05$\dagger\dagger$ &  & 2B & J1550-0538 & $<9$ \\
FBQS J1629+4007                & 0.272486 & 16 29 01.315 & +40 07 59.62 & 11.97 & b & 1.46 & pqr & 2B & J1623+3909$\dagger$ & 145 \\
2MASX J16332357+4718588        & 0.116054 & 16 33 23.585 & +47 18 58.96 & 62.63 & c & 2.22 & pqr & 2B & J1637+4717 & 163 \\
FBQS J1644+2619                & 0.145 & 16 44 42.536 & +26 19 13.19 & 87.53 & c & 2.65 & prs & 2B & J1642+2523 & 62 \\
B3 1702+457                    & 0.0604 & 17 03 30.379 & +45 40 47.09 & 115.44 & a & 2.17 & pqr & 2B & J1658+4737 & $<9$ \\
FBQS J1713+3523                & 0.083 & 17 13 04.476 & +35 23 33.43 & 11.13 & b & 1.05 &  & 2B & J1708+3346$\dagger$ & 138 \\
SDSS J172206.02+565451.6       & 0.425967 & 17 22 06.081 & +56 54 52.00 & 36.91 & c & 2.37 &  & 2B & J1722+5856 & $<9$ \\
\hline
\end{tabular}
}
\label{table:sample}
\begin{tabnote}
Col.~(1)~Source Name in NED; 
Col.~(2)~redshift; 
Col.~(3)~right ascension; 
Col.~(4)~declination; 
Col.~(5)~peak intensity in mJy~beam$^{-1}$ in VLA FIRST; 
Col.~(6)~parent sample; 
Col.~(7)~radio loudness from reference listed in Col.~(6), except for sources from \citet{Veron-Cetty:2001a} that includes no value for radio loudness.  Note that the definition of radio loudness differs from reference to reference.  We calculated radio loudness for sources in \citet{Veron-Cetty:2001a} by the ratio of FIRST 1.4~GHz radio to SDSS $g$-band PSF optical flux densities, according to \citet{Veron-Cetty:2001a}.  Sources denoted by $\dagger\dagger$ represent the use of $B$-band magnitudes listed in \citet{Veron-Cetty:2001a} because their SDSS results are not available;     
Col.~(8)~reference of previous VLBI observation; 
Col.~(9)~observation mode in the present study.  ''ND'' and ''2B'' represent nodding-style and VERA's dual-beam phase referencing, respectively.  ''1B'' represents an on-source observation with a single beam.  SDSS~J124634.64+023809.0 was mis-allocated in our observation; 
Col.~(10)~phase-referencing calibrator used in the present study.  Calibrators denoted by $\dagger$ represent non-detection in B-beam; 
Col.~(11)~correlated flux density in mJy measured in the present study.  
\\
References --- 
a: 	\citet{Veron-Cetty:2001a}	, 
b: 	\citet{Whalen:2006}	, 
c: 	\citet{Yuan:2008}	, 
d: 	\citet{Zhou:2006}	, 
e: 	\citet{Gu:2015}	, 
f: 	\citet{DAmmando:2013}	, 
g: 	\citet{Doi:2006}	, 
h: 	\citet{Abdo:2009b}	, 
i: 	\citet{Giroletti:2011}	, 
j: 	\citet{Doi:2013a}	, 
k: 	\citet{Doi:2015}	, 
l: 	\citet{Giroletti:2009}	, 
m: 	\citet{Fey:2000}	, 
n: 	\citet{DAmmando:2013a}	, 
o: 	\citet{Orienti:2015}	, 
p: 	\citet{Doi:2007}	, 
q: 	\citet{Gu:2010}	, 
r: 	\citet{Doi:2011a}	, 
s: 	\citet{Doi:2012}	
\end{tabnote}
\end{table*}


\section{Result and Discussion}\label{section:result}
Correlated flux densities averaged through baselines with detection are listed in the last column of Table~\ref{table:sample}; the table includes the references for sources for which previous VLBI studies exist.     
Upper limits of correlated flux density, ranging from $7$~mJy to $23$~mJy depending on the observing mode (Sect.~\ref{section:reduction}), were determined by $SNR=3$ on the most sensitive baseline during observations.  
Twelve out of 40 observed targets were detected with VERAs' baselines ranging from $\sim1000$~km to $\sim2300$~km, which imply a brightness temperature of the order of $10^7$~Kelvin or more for high-frequency radio emissions associated with these NLS1s.    
Even if we adopted a detection limit of $5\sigma$ instead of $3\sigma$, only [HB89]~1546+353, which is the weakest detection in our sample, would get into non-detection.

The two sources, FBQS~J114654.2+323652 and FBQS~J1713+3523, are for the first detected with very long baselines.   
FBQS~J114654.2+323652 was presumably detected with the MIZUSAWA--IRIKI baseline in the dual-beam phase reference mode; we also confirmed the fringe detection in a normal fringe fitting, without a calibrator, with a shorter integration (2~min).  Around this source, there is no strong source potentially confusing to our observation.  
FBQS~J1713+352 was positively detected with baselines of all the four antennas in fringe fitting without a calibrator (because its calibrator was not detected in B-beam).  Around this source, there is no strong source potentially confusing to our observation.  
Both the two sources are weak radio sources at 1.4~GHz in the FIRST (15.42~mJy and 11.24~mJy for FBQS~J114654.2+323652 (1994.4) and FBQS~J1713+352 (1994.5), respectively); on the other hand, they are relatively strong at 22~GHz in our observation (104~mJy and 138~mJy).  These sources should be inverted spectrum sources with $\alpha=+0.7$ and $\alpha=+0.9$, where $\alpha$ is the spectral index ($F_\nu \propto \nu^\alpha$), assuming that the 1.4~GHz FIRST (and the 22~GHz) fluxes do not vary significantly in  $\sim20$~years.    
Such a weak source showing an inverted spectrum has also been previously reported by \citet{Doi:2007} for FBQS~J1629+4007, which is also in the list of the present study and has been detected (11.95~mJy at 1.4~GHz (1994.6) and 145~mJy at 22~GHz).   
In the first place, relatively weak radio emissions at 1.4~GHz and a limited baseline sensitivity at 22~GHz were supposed to strongly bias VLBI detections toward inverted spectrum sources.  In fact, all the detected sources except for one (11/12) show flat or inverted spectra ($\alpha=-0.1$--$+0.9$) between 1.4--22~GHz, if variability in the time interval between measurements at 1.4 and 22~GHz is assumed insignificant.    
Although this is a sort of artificial effect, it is noteworthy that not so small fraction of NLS1 radio sources (12/40) have been detected at such a high frequency even with very long baselines.   

On the other hand, at least half of the sources indicate apparently steep spectra ($\alpha<-0.2$) between 1.4~GHz and 22~GHz on the basis of the FIRST flux densities and the VLBI upper limits.  Both the inverted and steep spectrum radio sources are included in the NLS1 population, as previously pointed out by several authors \citep{Doi:2007,Gu:2010,Doi:2011a}.  Particularly, $[$HB89$]$~1044+476 is the strongest radio source (767.44~mJy at 1.4~GHz in the FIRST (1997.2)) in our sample but not detected with the VLBI at 22~GHz ($<7$~mJy).  Only a weak radio emission (19.4~mJy) with a core-jet structure in the east-west direction has been previously found in VLBI images at 5~GHz while VLA images at 8.4 and 22~GHz show significantly large radio flux densities; a bulk of total flux must be resolved out with very long baselines and originate in extended components of a compact steep spectrum source \citep{Gu:2015}.  
B3~1702+457 is in the similar situation (118.64~mJy at FIRST 1.4~GHz (1997.2) and $<9$~mJy at VLBI 22~GHz); the previous VLBI studies revealed an extended radio structure with a steep spectrum for B3~1702+457 \citep{Doi:2007,Doi:2011a}.                     
Hence, our negative detections with a limited sensitivity at 22~GHz are genuine even for these sources relatively strong at 1.4~GHz.

Consequently, the systematic study by our VLBI detection survey has revealed that NLS1s are one of AGN subclasses that can possess compact radio components with high brightness temperatures ($\gtrsim10^7$~Kelvin) even at the high frequency (22~GHz).              
These compact components show inverted spectra in almost all the VLBI-detected cases, which account for a significant fraction in our radio-selected NLS1 sample.     
These properties may be related to NLS1s' blazer-like aspects such as gamma-ray detections in several NLS1s in contrast to normal broad-line Seyfert galaxies.

\begin{ack}
This study was partially supported by Grants-in-Aid for Scientific Research (B) (24340042, AD) and Grant-in-Aid for Scientific Research on Innovative Areas (26120537, AD) from the Japan Society for the Promotion of Science~(JSPS).   
We are grateful to all the staffs and students involved in the development and operation of the Japanese VLBI network~(JVN) and the VLBI Exploration of Radio Astrometry project~(VERA).  
The JVN project is led by the National Astronomical Observatory of Japan~(NAOJ), which is a branch of the National Institutes of Natural Sciences~(NINS), Yamaguchi University, Hokkaido University, Gifu University, Kagoshima University, Tsukuba University, Osaka Prefecture University, and Ibaraki University, in collaboration with the Geographical Survey Institute~(GSI), the Japan Aerospace Exploration Agency~(JAXA), and the National Institute of Information and Communications Technology~(NICT).  
The VERA is operated by the Mizusawa VLBI observatory, a branch of the National Astronomical Observatory of Japan.  
We used the US National Aeronautics and Space Administration's~(NASA) Astrophysics Data System~(ADS) abstract service and NASA/IPAC Extragalactic Database~(NED), which is operated by the Jet Propulsion Laboratory~(JPL).  
In addition, we used the Astronomical Image Processing System~(AIPS) software developed at the National Radio Astronomy Observatory~(NRAO), a facility of the US National Science Foundation operated under cooperative agreement by Associated Universities, Inc.  
\end{ack}


\begin{thebibliography}{43}
\expandafter\ifx\csname natexlab\endcsname\relax\def\natexlab#1{#1}\fi

\bibitem[{{Abdo} {et~al.}(2009{\natexlab{a}}){Abdo}, {Ackermann}, {Ajello},
  {Axelsson}, {Baldini}, {Ballet}, {Barbiellini}, {Bastieri}, {Battelino},
  {Baughman}, {Bechtol}, {Bellazzini}, {Bloom}, {Bonamente}, {Borgland},
  {Bregeon}, {Brez}, {Brigida}, {Bruel}, {Caliandro}, {Cameron}, {Caraveo},
  {Casandjian}, {Cavazzuti}, {Cecchi}, {Chekhtman}, {Cheung}, {Chiang},
  {Ciprini}, {Claus}, {Cohen-Tanugi}, {Collmar}, {Conrad}, {Costamante},
  {Dermer}, {de Angelis}, {de Palma}, {Digel}, {Silva}, {Drell}, {Dubois},
  {Dumora}, {Farnier}, {Favuzzi}, {Focke}, {Foschini}, {Frailis}, {Fuhrmann},
  {Fukazawa}, {Funk}, {Fusco}, {Gargano}, {Gehrels}, {Germani}, {Giebels},
  {Giglietto}, {Giordano}, {Giroletti}, {Glanzman}, {Grenier}, {Grondin},
  {Grove}, {Guillemot}, {Guiriec}, {Hanabata}, {Harding}, {Hartman},
  {Hayashida}, {Hays}, {Hughes}, {J{\'o}hannesson}, {Johnson}, {Johnson},
  {Johnson}, {Kamae}, {Katagiri}, {Kataoka}, {Kerr}, {Kn{\"o}dlseder}, {Kuehn},
  {Kuss}, {Lande}, {Latronico}, {Lemoine-Goumard}, {Longo}, {Loparco}, {Lott},
  {Lovellette}, {Lubrano}, {Madejski}, {Makeev}, {Max-Moerbeck}, {Mazziotta},
  {McConville}, {McEnery}, {Meurer}, {Michelson}, {Mitthumsiri}, {Mizuno},
  {Monte}, {Monzani}, {Morselli}, {Moskalenko}, {Murgia}, {Nolan}, {Norris},
  {Nuss}, {Ohsugi}, {Omodei}, {Orlando}, {Ormes}, {Paneque}, {Panetta},
  {Parent}, {Pavlidou}, {Pearson}, {Pepe}, {Pesce-Rollins}, {Piron}, {Porter},
  {Rain{\`o}}, {Rando}, {Razzano}, {Readhead}, {Reimer}, {Reimer}, {Reposeur},
  {Richards}, {Ritz}, {Rodriguez}, {Romani}, {Ryde}, {Sadrozinski}, {Sambruna},
  {Sanchez}, {Sander}, {Parkinson}, {Scargle}, {Schalk}, {Sgr{\`o}}, {Smith},
  {Spandre}, {Spinelli}, {Starck}, {Stevenson}, {Strickman}, {Suson},
  {Tagliaferri}, {Takahashi}, {Tanaka}, {Thayer}, {Thompson}, {Tibaldo},
  {Tibolla}, {Torres}, {Tosti}, {Tramacere}, {Uchiyama}, {Usher}, {Vilchez},
  {Vitale}, {Waite}, {Winer}, {Wood}, {Ylinen}, {Zensus}, {Ziegler}, {Fermi/LAT
  Collaboration}, {Ghisellini}, {Maraschi}, {Tavecchio}, \&
  {Angelakis}}]{Abdo:2009a}
{Abdo}, A.~A., {et~al.} 2009{\natexlab{a}}, \apj, 699, 976

\bibitem[{{Abdo} {et~al.}(2009{\natexlab{b}}){Abdo}, {Ackermann}, {Ajello},
  {Axelsson}, {Baldini}, {Ballet}, {Barbiellini}, {Bastieri}, {Baughman},
  {Bechtol}, \& et~al.}]{Abdo:2009b}
---. 2009{\natexlab{b}}, \apj, 707, 727

\bibitem[{{Abdo} {et~al.}(2009{\natexlab{c}}){Abdo}, {Ackermann}, {Ajello},
  {Baldini}, {Ballet}, {Barbiellini}, {Bastieri}, {Bechtol}, {Bellazzini},
  {Berenji}, {Bloom}, {Bonamente}, {Borgland}, {Bregeon}, {Brez}, {Brigida},
  {Bruel}, {Burnett}, {Caliandro}, {Cameron}, {Caraveo}, {Casandjian},
  {Cecchi}, {{\c C}elik}, {Chekhtman}, {Cheung}, {Chiang}, {Ciprini}, {Claus},
  {Cohen-Tanugi}, {Conrad}, {Cutini}, {Dermer}, {de Palma}, {Silva}, {Drell},
  {Dubois}, {Dumora}, {Farnier}, {Favuzzi}, {Fegan}, {Focke}, {Foschini},
  {Frailis}, {Fukazawa}, {Fusco}, {Gargano}, {Gehrels}, {Germani}, {Giebels},
  {Giglietto}, {Giordano}, {Giroletti}, {Glanzman}, {Godfrey}, {Grenier},
  {Grove}, {Guillemot}, {Guiriec}, {Hayashida}, {Hays}, {Horan}, {Hughes},
  {J{\'o}hannesson}, {Johnson}, {Johnson}, {Kadler}, {Kamae}, {Katagiri},
  {Kataoka}, {Kerr}, {Kn{\"o}dlseder}, {Kuss}, {Lande}, {Latronico}, {Longo},
  {Loparco}, {Lott}, {Lovellette}, {Lubrano}, {Makeev}, {Mazziotta},
  {McConville}, {McEnery}, {Meurer}, {Michelson}, {Mitthumsiri}, {Mizuno},
  {Monte}, {Monzani}, {Morselli}, {Moskalenko}, {Murgia}, {Nolan}, {Norris},
  {Nuss}, {Ohsugi}, {Omodei}, {Orlando}, {Ormes}, {Pelassa}, {Pepe}, {Persic},
  {Pesce-Rollins}, {Piron}, {Porter}, {Rain{\`o}}, {Rando}, {Razzano},
  {Rochester}, {Rodriguez}, {Ryde}, {Sadrozinski}, {Sambruna}, {Sander}, {Saz
  Parkinson}, {Scargle}, {Sgr{\`o}}, {Smith}, {Spandre}, {Spinelli},
  {Strickman}, {Suson}, {Tagliaferri}, {Takahashi}, {Takahashi}, {Tanaka},
  {Thayer}, {Thayer}, {Thompson}, {Tibaldo}, {Tibolla}, {Torres}, {Tosti},
  {Tramacere}, {Uchiyama}, {Usher}, {Vasileiou}, {Vilchez}, {Vitale}, {Waite},
  {Wang}, {Winer}, {Wood}, {Ylinen}, {Ziegler}, {Fermi/LAT Collaboration},
  {Ghisellini}, {Maraschi}, \& {Tavecchio}}]{Abdo:2009}
---. 2009{\natexlab{c}}, \apjl, 707, L142

\bibitem[{{Angelakis} {et~al.}(2015){Angelakis}, {Fuhrmann}, {Marchili},
  {Foschini}, {Myserlis}, {Karamanavis}, {Komossa}, {Blinov}, {Krichbaum},
  {Sievers}, {Ungerechts}, \& {Zensus}}]{Angelakis:2015}
{Angelakis}, E., {et~al.} 2015, \aap, 575, A55

\bibitem[{{Ant{\'o}n} {et~al.}(2008){Ant{\'o}n}, {Browne}, \&
  {March{\~a}}}]{Anton:2008}
{Ant{\'o}n}, S., {Browne}, I.~W.~A., \& {March{\~a}}, M.~J. 2008, \aap, 490,
  583

\bibitem[{{Becker} {et~al.}(1995){Becker}, {White}, \& {Helfand}}]{Becker:1995}
{Becker}, R.~H., {White}, R.~L., \& {Helfand}, D.~J. 1995, \apj, 450, 559

\bibitem[{{Condon} {et~al.}(1998){Condon}, {Cotton}, {Greisen}, {Yin},
  {Perley}, {Taylor}, \& {Broderick}}]{Condon:1998}
{Condon}, J.~J., {Cotton}, W.~D., {Greisen}, E.~W., {Yin}, Q.~F., {Perley},
  R.~A., {Taylor}, G.~B., \& {Broderick}, J.~J. 1998, \aj, 115, 1693

\bibitem[{{D'Ammando} {et~al.}(2015){D'Ammando}, {Orienti}, {Larsson}, \&
  {Giroletti}}]{DAmmando:2015a}
{D'Ammando}, F., {Orienti}, M., {Larsson}, J., \& {Giroletti}, M. 2015, \mnras,
  452, 520

\bibitem[{{D'Ammando} {et~al.}(2012){D'Ammando}, {Orienti}, {Finke}, {Raiteri},
  {Angelakis}, {Fuhrmann}, {Giroletti}, {Hovatta}, {Max-Moerbeck}, {Perkins},
  {Readhead}, {Richards}, {Stawarz}, \& {Donato}}]{DAmmando:2012}
{D'Ammando}, F., {et~al.} 2012, \mnras, 426, 317

\bibitem[{{D'Ammando} {et~al.}(2013{\natexlab{a}}){D'Ammando}, {Orienti},
  {Finke}, {Raiteri}, {Angelakis}, {Fuhrmann}, {Giroletti}, {Hovatta},
  {Karamanavis}, {Max-Moerbeck}, {Myserlis}, {Readhead}, \&
  {Richards}}]{DAmmando:2013}
---. 2013{\natexlab{a}}, \mnras, 436, 191

\bibitem[{{D'Ammando} {et~al.}(2013{\natexlab{b}}){D'Ammando}, {Orienti},
  {Doi}, {Giroletti}, {Dallacasa}, {Hovatta}, {Drake}, {Max-Moerbeck},
  {Readhead}, \& {Richards}}]{DAmmando:2013a}
---. 2013{\natexlab{b}}, \mnras, 433, 952

\bibitem[{{Doi} {et~al.}(2013){Doi}, {Asada}, {Fujisawa}, {Nagai}, {Hagiwara},
  {Wajima}, \& {Inoue}}]{Doi:2013a}
{Doi}, A., {Asada}, K., {Fujisawa}, K., {Nagai}, H., {Hagiwara}, Y., {Wajima},
  K., \& {Inoue}, M. 2013, \apj, 765, 69

\bibitem[{{Doi} {et~al.}(2011){Doi}, {Asada}, \& {Nagai}}]{Doi:2011a}
{Doi}, A., {Asada}, K., \& {Nagai}, H. 2011, \apj, 738, 126

\bibitem[{{Doi} {et~al.}(2006){Doi}, {Nagai}, {Asada}, {Kameno}, {Wajima}, \&
  {Inoue}}]{Doi:2006}
{Doi}, A., {Nagai}, H., {Asada}, K., {Kameno}, S., {Wajima}, K., \& {Inoue}, M.
  2006, \pasj, 58, 829

\bibitem[{{Doi} {et~al.}(2012){Doi}, {Nagira}, {Kawakatu}, {Kino}, {Nagai}, \&
  {Asada}}]{Doi:2012}
{Doi}, A., {Nagira}, H., {Kawakatu}, N., {Kino}, M., {Nagai}, H., \& {Asada},
  K. 2012, \apj, 760, 41

\bibitem[{{Doi} {et~al.}(2015){Doi}, {Wajima}, {Hagiwara}, \&
  {Inoue}}]{Doi:2015}
{Doi}, A., {Wajima}, K., {Hagiwara}, Y., \& {Inoue}, M. 2015, \apjl, 798, L30

\bibitem[{{Doi} {et~al.}(2007){Doi}, {Fujisawa}, {Inoue}, {Wajima}, {Nagai},
  {Harada}, {Suematsu}, {Habe}, {Honma}, {Kawaguchi}, {Kawai}, {Kobayashi},
  {Koyama}, {Kuboki}, {Murata}, {Omodaka}, {Sorai}, {Sudou}, {Takaba},
  {Takashima}, {Takeda}, {Tamura}, \& {Wakamatsu}}]{Doi:2007}
{Doi}, A., {et~al.} 2007, \pasj, 59, 703

\bibitem[{{Fey} \& {Charlot}(2000)}]{Fey:2000}
{Fey}, A.~L., \& {Charlot}, P. 2000, \apjs, 128, 17

\bibitem[{{Foschini}(2011)}]{Foschini:2011}
{Foschini}, L. 2011, in Narrow-Line Seyfert 1 Galaxies and their Place in the
  Universe

\bibitem[{{Fujisawa et al.}(submitted)}]{Fujisawa:2016}
{Fujisawa et al.}, K. submitted, \pasj

\bibitem[{{Giroletti} \& {Panessa}(2009)}]{Giroletti:2009}
{Giroletti}, M., \& {Panessa}, F. 2009, \apjl, 706, L260

\bibitem[{{Giroletti} {et~al.}(2005){Giroletti}, {Taylor}, \&
  {Giovannini}}]{Giroletti:2005}
{Giroletti}, M., {Taylor}, G.~B., \& {Giovannini}, G. 2005, \apj, 622, 178

\bibitem[{{Giroletti} {et~al.}(2011){Giroletti}, {Paragi}, {Bignall}, {Doi},
  {Foschini}, {Gab{\'a}nyi}, {Reynolds}, {Blanchard}, {Campbell}, {Colomer},
  {Hong}, {Kadler}, {Kino}, {van Langevelde}, {Nagai}, {Phillips}, {Sekido},
  {Szomoru}, \& {Tzioumis}}]{Giroletti:2011}
{Giroletti}, M., {et~al.} 2011, \aap, 528, L11

\bibitem[{{Gliozzi} {et~al.}(2010){Gliozzi}, {Papadakis}, {Grupe}, {Brinkmann},
  {Raeth}, \& {Kedziora-Chudczer}}]{Gliozzi:2010}
{Gliozzi}, M., {Papadakis}, I.~E., {Grupe}, D., {Brinkmann}, W.~P., {Raeth},
  C., \& {Kedziora-Chudczer}, L. 2010, \apj, 717, 1243

\bibitem[{{Greisen}(2003)}]{Greisen:2003}
{Greisen}, E.~W. 2003, Information Handling in Astronomy - Historical Vistas,
  285, 109

\bibitem[{{Grupe} {et~al.}(2000){Grupe}, {Leighly}, {Thomas}, \&
  {Laurent-Muehleisen}}]{Grupe:2000}
{Grupe}, D., {Leighly}, K.~M., {Thomas}, H.-C., \& {Laurent-Muehleisen}, S.~A.
  2000, \aap, 356, 11

\bibitem[{{Gu} \& {Chen}(2010)}]{Gu:2010}
{Gu}, M., \& {Chen}, Y. 2010, \aj, 139, 2612

\bibitem[{{Gu} {et~al.}(2015){Gu}, {Chen}, {Komossa}, {Yuan}, {Shen}, {Wajima},
  {Zhou}, \& {Zensus}}]{Gu:2015}
{Gu}, M., {Chen}, Y., {Komossa}, S., {Yuan}, W., {Shen}, Z., {Wajima}, K.,
  {Zhou}, H., \& {Zensus}, J.~A. 2015, \apjs, 221, 3

\bibitem[{{Kobayashi} {et~al.}(2003){Kobayashi}, {Sasao}, {Kawaguchi},
  {Manabe}, {Omodaka}, {Kameya}, {Shibata}, {Miyaji}, {Honma}, {Tamura},
  {Hirota}, {Kuji}, {Horiai}, {Sakai}, {Sato}, {Iwadate}, {Kanya}, {Ujihara},
  {Jike}, {Fujii}, {Motiduki}, {Oyama}, {Kurayama}, {Kamohara}, {Suda}, \&
  {Kasuga}}]{Kobayashi:2003}
{Kobayashi}, H., {et~al.} 2003, in Astronomical Society of the Pacific
  Conference Series, Vol. 306, New technologies in VLBI, ed. Y.~C. {Minh}, 367

\bibitem[{{Komossa} {et~al.}(2006){Komossa}, {Voges}, {Xu}, {Mathur}, {Adorf},
  {Lemson}, {Duschl}, \& {Grupe}}]{Komossa:2006}
{Komossa}, S., {Voges}, W., {Xu}, D., {Mathur}, S., {Adorf}, H.-M., {Lemson},
  G., {Duschl}, W.~J., \& {Grupe}, D. 2006, \aj, 132, 531

\bibitem[{{Kono} {et~al.}(2012){Kono}, {Oyama}, {Kawaguchi}, {Suzuki},
  {Fujisawa}, {Takaba}, {Sorai}, {Sekido}, {Kurihara}, {Murata}, \&
  {Uose}}]{Kono:2012}
{Kono}, Y., {et~al.} 2012, in Seventh General Meeting (GM2012) of the
  international VLBI Service for Geodesy and Astrometry (IVS), held in Madrid,
  Spain, March 4-9, 2012, Eds: D. Behrend and K.D. Baver, National Aeronautics
  and Space Administration, ed. D.~{Behrend} \& K.~D. {Baver}

\bibitem[{{Moran}(2000)}]{Moran:2000}
{Moran}, E.~C. 2000, \nar, 44, 527

\bibitem[{{Orienti} {et~al.}(2015){Orienti}, {D'Ammando}, {Larsson}, {Finke},
  {Giroletti}, {Dallacasa}, {Isacsson}, \& {Stoby Hoglund}}]{Orienti:2015}
{Orienti}, M., {D'Ammando}, F., {Larsson}, J., {Finke}, J., {Giroletti}, M.,
  {Dallacasa}, D., {Isacsson}, T., \& {Stoby Hoglund}, J. 2015, \mnras, 453,
  4037

\bibitem[{{Oshlack} {et~al.}(2001){Oshlack}, {Webster}, \&
  {Whiting}}]{Oshlack:2001}
{Oshlack}, A.~Y.~K.~N., {Webster}, R.~L., \& {Whiting}, M.~T. 2001, \apj, 558,
  578

\bibitem[{{Oyama} {et~al.}(2012){Oyama}, {Kono}, {Suzuki}, {Mizuno},
  {Bushimata}, {Jike}, {Kawaguchi}, {Kobayashi}, \& {Kimura}}]{Oyama:2012}
{Oyama}, T., {et~al.} 2012, in Seventh General Meeting (GM2012) of the
  international VLBI Service for Geodesy and Astrometry (IVS), held in Madrid,
  Spain, March 4-9, 2012, Eds: D. Behrend and K.D. Baver, National Aeronautics
  and Space Administration, p. 91-95, ed. D.~{Behrend} \& K.~D. {Baver}, 91--95

\bibitem[{{Richards} \& {Lister}(2015)}]{Richards:2015}
{Richards}, J.~L., \& {Lister}, M.~L. 2015, \apjl, 800, L8

\bibitem[{{Siebert} {et~al.}(1999){Siebert}, {Leighly}, {Laurent-Muehleisen},
  {Brinkmann}, {Boller}, \& {Matsuoka}}]{Siebert:1999}
{Siebert}, J., {Leighly}, K.~M., {Laurent-Muehleisen}, S.~A., {Brinkmann}, W.,
  {Boller}, T., \& {Matsuoka}, M. 1999, \aap, 348, 678

\bibitem[{{Ulvestad} {et~al.}(1995){Ulvestad}, {Antonucci}, \&
  {Goodrich}}]{Ulvestad:1995}
{Ulvestad}, J.~S., {Antonucci}, R.~R.~J., \& {Goodrich}, R.~W. 1995, \aj, 109,
  81

\bibitem[{{V{\'e}ron-Cetty} {et~al.}(2001){V{\'e}ron-Cetty}, {V{\'e}ron}, \&
  {Gon{\c c}alves}}]{Veron-Cetty:2001a}
{V{\'e}ron-Cetty}, M.-P., {V{\'e}ron}, P., \& {Gon{\c c}alves}, A.~C. 2001,
  \aap, 372, 730

\bibitem[{{Wajima} {et~al.}(2014){Wajima}, {Fujisawa}, {Hayashida}, {Isobe},
  {Ishida}, \& {Yonekura}}]{Wajima:2014}
{Wajima}, K., {Fujisawa}, K., {Hayashida}, M., {Isobe}, N., {Ishida}, T., \&
  {Yonekura}, Y. 2014, \apj, 781, 75

\bibitem[{{Whalen} {et~al.}(2006){Whalen}, {Laurent-Muehleisen}, {Moran}, \&
  {Becker}}]{Whalen:2006}
{Whalen}, D.~J., {Laurent-Muehleisen}, S.~A., {Moran}, E.~C., \& {Becker},
  R.~H. 2006, \aj, 131, 1948

\bibitem[{{Yuan} {et~al.}(2008){Yuan}, {Zhou}, {Komossa}, {Dong}, {Wang}, {Lu},
  \& {Bai}}]{Yuan:2008}
{Yuan}, W., {Zhou}, H.~Y., {Komossa}, S., {Dong}, X.~B., {Wang}, T.~G., {Lu},
  H.~L., \& {Bai}, J.~M. 2008, \apj, 685, 801

\bibitem[{{Zhou} {et~al.}(2006){Zhou}, {Wang}, {Yuan}, {Lu}, {Dong}, {Wang}, \&
  {Lu}}]{Zhou:2006}
{Zhou}, H., {Wang}, T., {Yuan}, W., {Lu}, H., {Dong}, X., {Wang}, J., \& {Lu},
  Y. 2006, \apjs, 166, 128

\end{thebibliography}


\end{document}